\documentclass{JINST}
\usepackage{amsmath,amssymb,fancyhdr,textcomp,cite}
\usepackage[pdftex]{graphicx}

\title{Cryogenic CMOS Cameras for High Voltage Monitoring in Liquid Argon}

\author{ N. McConkey\thanks{Corresponding author.}, N. Spooner, M. Thiesse, M. Wallbank and T. K. Warburton\\
University of Sheffield, \\
Department of Physics and Astronomy, Hicks Building, Hounsfield Road, Sheffield, S3 7RH, UK \\
E-mail: \email{n.mcconkey@sheffield.ac.uk}}

\abstract{The prevalent use of large volume liquid argon detectors strongly motivates the development of novel readout and monitoring technology which functions at cryogenic temperatures.  This paper presents the development of a cryogenic CMOS camera system suitable for use inside a large volume liquid argon detector for online monitoring purposes.  The characterisation of the system is described in detail.  The reliability of such a camera system has been demonstrated over several months, and recent data from operation within the liquid argon region of the DUNE 35t cryostat is presented.  The cameras were used to monitor for high voltage breakdown inside the cryostat, with capability to observe breakdown of a liquid argon time projection chamber in situ.  They were also used for detector monitoring, especially of components during cooldown.}


\begin{document}

\section{Introduction}

Liquid argon (LAr) and xenon (LXe) are popular choices of medium for detectors in rare event physics such as neutrino physics \cite{LAroverview} and dark matter searches \cite{cryodarkmatter}. This is due partly to the excellent charge transport and scintillation properties of these materials, along with their high density. The low boiling points of 88K and 162K, respectively, of these liquids then strongly motivates development of various new technologies that can operate successfully in cryogenic conditions.

The work presented here concerns the use of Complementary Metal-Oxide Semiconductor (CMOS) cameras at cryogenic temperature within LAr, with a view to using this technology for optical readout in LAr and LXe detectors. Charge-coupled device (CCD) and CMOS cameras have previously been used to monitor LAr detectors \cite{BERNcam,LAPD,Liverpool,breakdown2,Weizmannbubbles}. However, this has been either from outside the cryostat, through a view port, or by incorporation of an enclosure with raised temperature. The development of devices that function directly inside the cryostat liquid gives many advantages over such externally located or heated cameras. This includes improved applicability to larger scale detectors but also potential for use in readout of secondary scintillation light \cite{electroluminescence}. The development of such devices also holds potential to avoid liquid bubbling or other disturbances that can occur when using devices that require heating.

One application of cameras operated within LAr is the potential to monitor for high voltage (HV) discharge problems within a cryostat. Successful operation with HV in LAr, for example as the drift voltage for a Time Projection Chamber (TPC), has recently become an area for concern. In the 1960s, the breakdown field in LAr was measured in \cite{breakdown1} to be 1.4$\,$MVcm$^{-1}$. However, recent research has demonstrated that at the length scale and purity level desired in a TPC for neutrino or dark matter physics, the breakdown field can be much lower than this, at fields of around 40$\,$kVcm$^{-1}$ \cite{breakdown2}. For large volume TPCs, such as those proposed for the Deep Underground Neutrino Experiment (DUNE), this may imply significant extra constraints on the design. For instance, the drift field required for DUNE is 500$\,$Vcm$^{-1}$, implying a nominal drift voltage as high as -190 kV \cite{DUNE}. Renewed study of HV breakdown has thus become a key topic in order to mitigate against the possibility of discharge in such situations.

Within the wider context of developing in-liquid optical readout for liquid noble gas detectors, the focus of this paper is on development of a new online monitoring system using cryogenic CMOS cameras for observing such HV breakdown in situ, in an operational TPC. Specifically the work concerns design and operation of such a system for the DUNE 35-ton prototype experiment. This detector, run at Fermilab in early 2016, was made up of prototype components for the DUNE far detector \cite{DUNE}. The camera system was designed to visually monitor the TPC and cryogenic support components from inside the argon-filled cryostat whilst the TPC was running, particularly for any high field areas. The cameras are sensitive to visible light, and can therefore observe sparks and coronae inside the cryostat. The motivation here was thus to enable location of any parts of the detector that might undergo HV breakdown, to troubleshoot any difficulties with the HV which are not possible to diagnose using other methods such as power supply current monitoring, and hence also aid the technology development for the next iteration of TPC design for DUNE.
\section{Cryogenic CMOS camera testing}

\subsection{Camera selection}

The primary basis for selection of the cameras was the ability to operate at cryogenic temperatures with sufficient sensitivity to detect light pulses from sparks.  Cameras based on CMOS technology were tested, since CMOS sensors themselves have been shown to operate well at cryogenic temperatures due to the semiconductor properties changing favourably.  Relative to room temperature a CMOS chip has lower leakage current, improved mobility and lower thermal noise at temperatures around 100K \cite{thermalnoise1}, \cite{thermalnoise2}.

The use of CMOS based cameras is widespread, and there are a plethora of camera systems designed to work at temperatures down to -40$\,^\circ$C (233$\,$K), but there are few systems designed to work below that, especially not commercially.   The initial selection of cameras was based on the requirement for a simple and compact system, at low cost. Cameras were selected which were manufacturer rated to -40$\,^\circ$C.  A large variety of these are sold as car-reversing cameras.  

A cryogenic shock test showed that many of the cameras failed when cooled to 77K, and others experienced a significant degradation of the signal.  This is due to many electronic components not functioning within design specification at cryogenic temperatures \cite{cryocompfail}.  This leads to non-functioning circuits for the operation of the CMOS sensor, and hence failure at cryogenic temperatures.  

The CMOS camera which is characterised in this paper, a \textit{Floureon} car reversing camera, experienced minimal change in resolution when cryogenically shock tested.  It has a simple internal circuit which is composed of resistors, capacitors, a crystal oscillator, and a flash memory chip.  These types of components have been studied in previous works and certain devices have been characterised and demonstrated to work successfully at cryogenic temperatures \cite{crystalosc}, \cite{flashmem}.

\subsection{Camera characterisation}\label{campar}

The key performance parameters of the selected camera type were measured initially at room temperature, and verified at cryogenic temperatures, to characterise their operation whilst immersed in liquid argon.  These are summarised in table \ref{partable}. This measurement programme is described below.

\begin{table}[h!]
\centering
\caption{Camera parameters summary table\label{partable}}
\begin{tabular}{|l l l|}
\hline
Temperature (K) & 290 & 77 \\
\hline
Pixels & 712$\times$486 & 712$\times$486\\
Frame rate & 50$\,$Hz & 50$\,$Hz \\
Viewing angle & 107$^\circ$ & - \\
Resolution at 10$\,$mm & 2$\pm$0.5$\,$mm & 1.5$\pm$0.5$\,$mm\\
Resolution at 2$\,$m & 50$\pm$5$\,$mm & - \\
Minimum measurable light pulse width & 20$\,$ns & 20$\,$ns \\ \hline
\end{tabular}
\end{table}

\begin{figure}[b]
\centering
\includegraphics[width=0.7\textwidth]{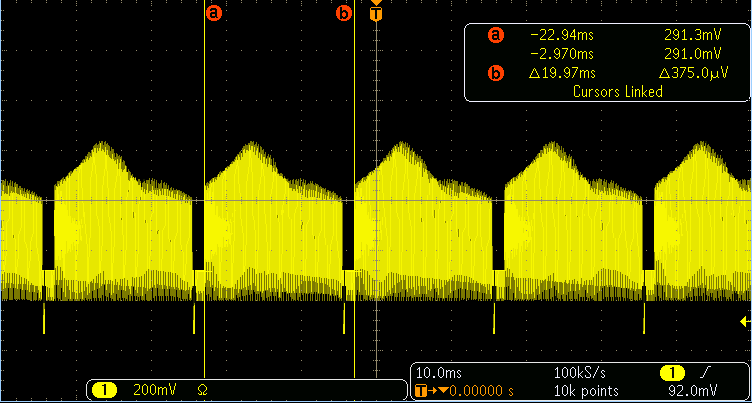}
\caption{The oscilloscope trace from the output of a CMOS camera.  The period of 20ms, as shown by the cursors corresponds to the camera frame rate of 50$\,$FPS.\label{scopetrace}}
\end{figure}

The device has a CMOS image sensor with 712 x 486 pixels, rolling shutter frame rate of 50 frames per second (FPS), and a PAL formatted video signal.  The frame rate was determined by observing the raw composite video signal using an oscilloscope. Figure \ref{scopetrace} shows an example trace showing the period of 20$\,$ms which corresponds to the frame rate. The frame rate is determined by the crystal oscillator component in the camera circuit.  It is a component which has typically been shown to vary as a function of temperature \cite{crystalosc},\cite{crystaldatasheet}.  A typical variation is of the order of 1$\times$10$^{-8}$Hz per $^\circ$C, which for the 29.5$\,$MHz crystal in the camera circuit, operating at 77K gives an oscillation difference which is an insignificant variation in oscillation frequency, within the device specification tolerance.  With the camera immersed in liquid nitrogen, the measured frame rate of 50$\,$Hz remained constant.  

\begin{figure}
\centering
\includegraphics[width=0.55\textwidth]{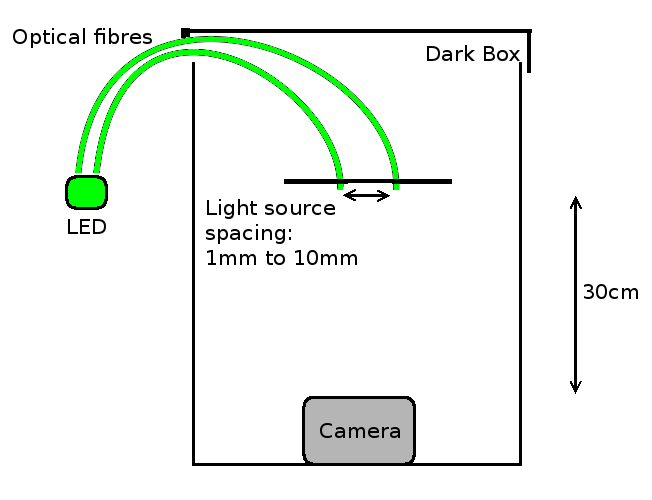}
\caption{A schematic of the test setup used for measuring the resolution at short distances at room and cryogenic temperatures. \label{ressetup}}
\end{figure}

\begin{figure}
\centering
\includegraphics[width=0.7\textwidth]{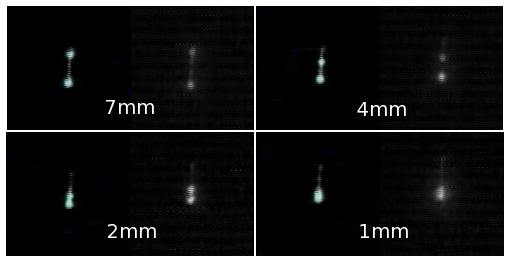}
\caption{Photographs comparing the camera resolution in warm (left) and cold (right) at a focal length of 0.1$\,$m, with optical fibre spacings of 7$\,$mm, 4$\,$mm, 2$\,$mm and 1$\,$mm.\label{rescryocompare}}
\end{figure}

The resolution of the cameras is critical to their performance.  Comparative measurements of the resolving power were made in liquid nitrogen and air using two optical fibres and a green LED in a dark box, with an LED to camera spacing of 0.1$\,$m, as shown in the schematic in figure \ref{ressetup}.  The optical fiber spacing was varied, and the resolution determined. Figure \ref{rescryocompare} shows snapshots from this, with varied optical fibre spacing.  The image on the left in each case is the room temperature measurement, taken in air, and the image on the right is the cryogenic measurement, taken in liquid nitrogen. The resolution is shown to be unaffected by cryogenic temperatures, within systematic errors.  The main uncertainty on this measurement is the variation in the spread of the transmitted light - the lower refractive index of air causes the light to be more diffuse, and hence gives poorer resolution. 

\begin{figure}
\centering
\includegraphics[width=0.7\textwidth]{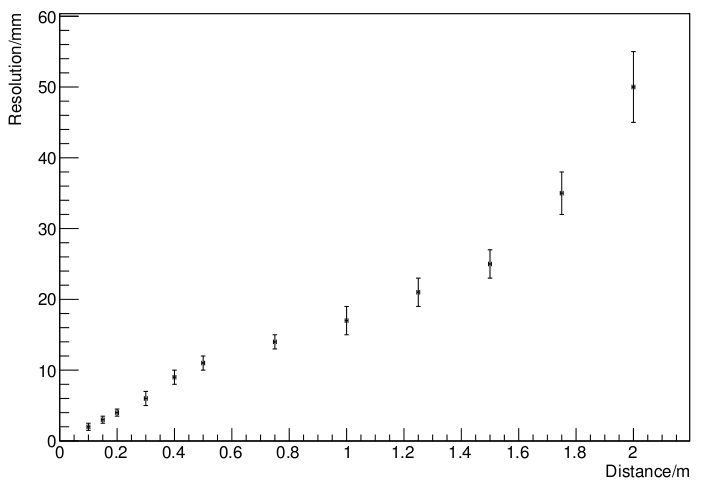}
\caption{The variation in the camera resolution as a function of distance, measured at room temperature using pattern recognition.\label{resolutionplot}}
\end{figure}

The resolution variation was additionally measured over longer distances, similar to those appropriate in the context of detector monitoring.  This was a basic measurement made in air only, using pattern recognition to determine the ability of the camera to resolve patterns of varied size. The resolution is taken as the smallest circle size resolvable at a separation distance equal to its diameter.  The resolution varies as a function of distance from the camera, as is shown in figure \ref{resolutionplot}.  

The timing sensitivity of the camera response determines its applicability to measuring discharges inside a detector.  For this camera system, the sensitivity depends both on the software trigger and the device itself. The timing sensitivity of the camera system was measured using an LED in a dark box, connected to a pulse generator which produced variable width pulses.  The camera was able to trigger on pulses with width 20$\,$ns. The rolling shutter nature of the camera means that each row of the CMOS chip is exposed for the majority of the 20$\,$ms in which it is not being read out. 

The software trigger for the camera system is sensitive to the difference between pixels in two consecutive frames.  The percentage of pixel difference between two frames gives the trigger threshold.  With respect to the measurement above, the trigger sensitivity for the fast pulses was limited by the spacial extent of the light flashes being produced.  It is anticipated that faster pulses could be detected if the image response had a higher light intensity, over a larger spacial extent.  

In order to test the response of the camera to sparks, high voltage was applied across a printed circuit board (PCB) in liquid argon, until breakdown was observed.  The temporal extent of the discharge was between 40 and 60$\,$ms, and camera response showed a localised spark over multiple frames of exposure.  The cameras were able to successfully detect and trigger on these sparks.

\begin{figure}
\centering
\includegraphics[width=0.7\textwidth]{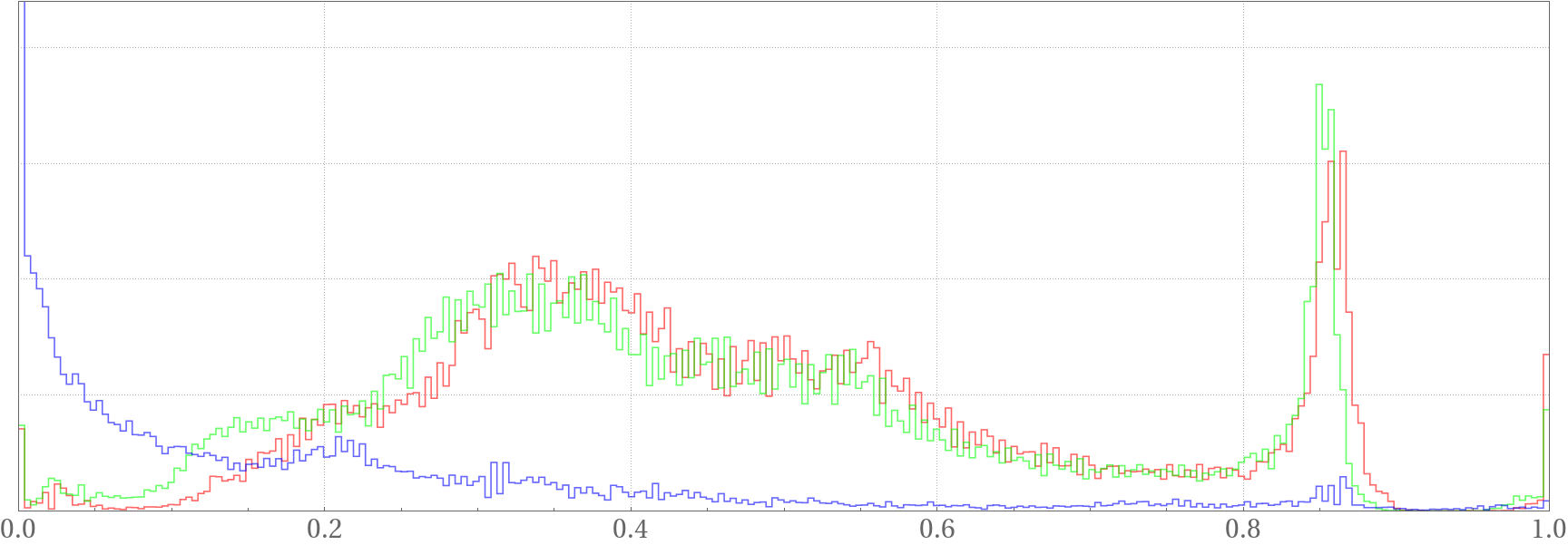}
\includegraphics[width=0.7\textwidth]{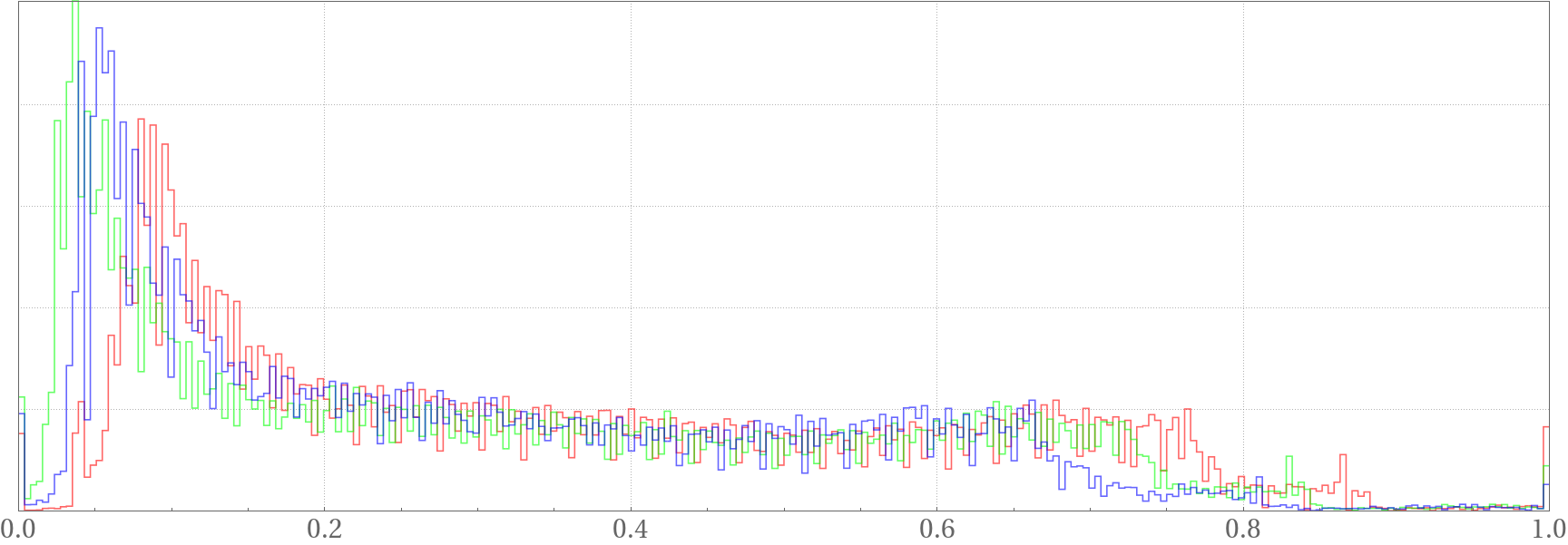}
\caption{RGB histograms of images taken before (top) and after (bottom) cooling the camera from room-temperature to LAr temperature. Red, green, and blue histograms are overlaid in each plot, showing the relative fraction of total pixels containing each colour shown on the horizontal axis.\label{lumachroma}}
\end{figure}

A notable behavioural change with regards to the operation of the selected camera type is the chrominance output of the video signal. The devices normally output a colour video signal at room temperature, but a monochrome signal at cryogenic temperatures. The authors suggest this difference in functionality is due to an unidentified partial failure of the on-board encoding circuits in the cold. The digitized signal from the silicon sensors is converted into an analogue \textit{PAL} signal for transmission. The `full colour' chrominance signal is split into three separate signals to be translated: Y'(black and white component), U (blue-luma), and V (red-luma) \cite{lumachroma}. The luminance and Y' encoding appears functional at cryogenic temperatures, but the U and V signals do not, hence a monochrome image is output in the cold. Loss of colour encoding is shown in the bottom plot of figure \ref{lumachroma} by the equalising of colour ratios amongst the image pixels.

\subsection{Heating and reliability}

The cameras characterised in the above section operate at cryogenic temperatures without need for heating or insulation.  This section describes the reliability of operation, and discusses steps taken to mitigate camera failure modes which could occur in longer term operation of such devices.

\begin{figure}
	\minipage{0.5\textwidth}
		\includegraphics[width=\linewidth]{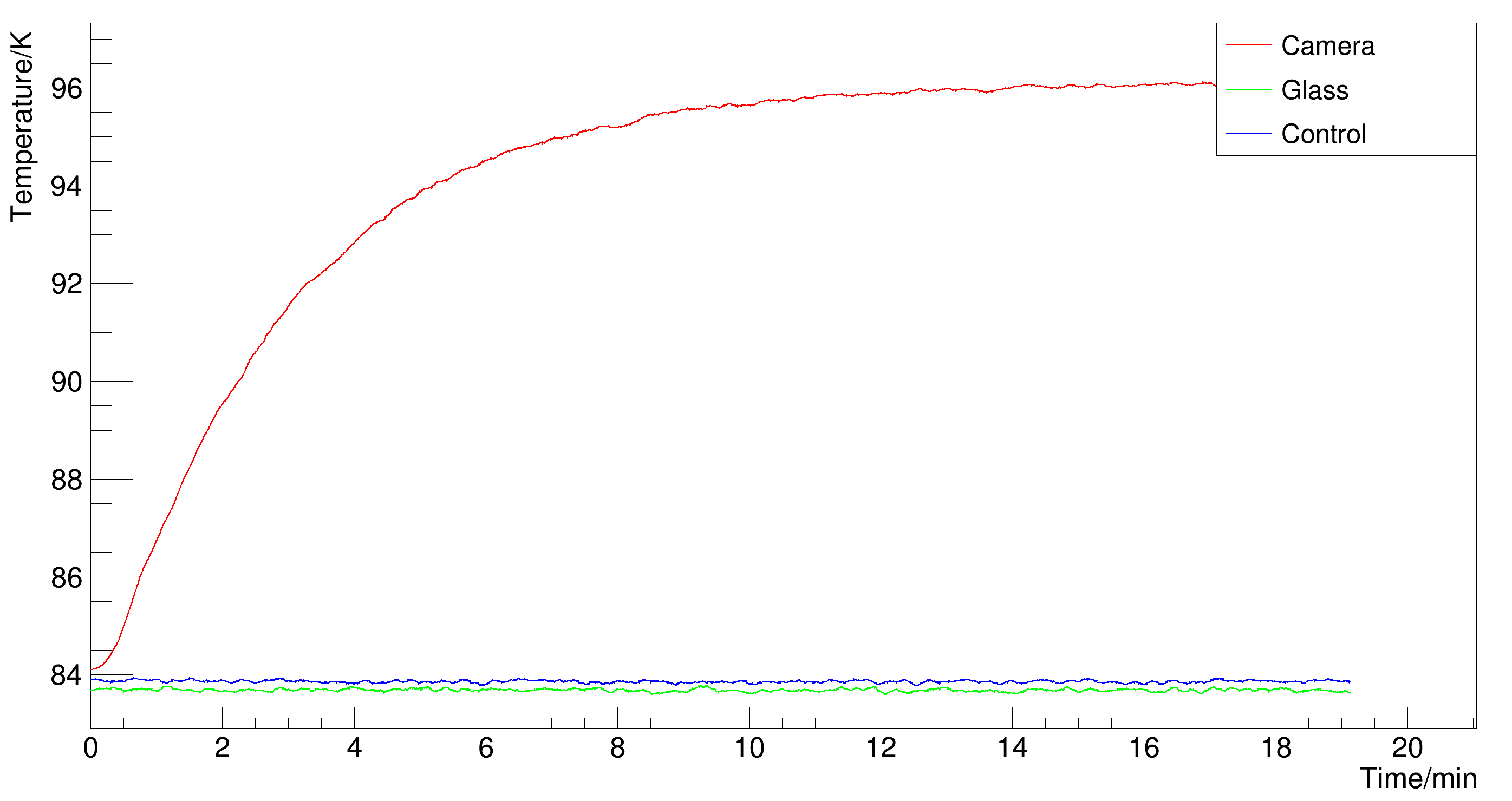}
	\endminipage\hfill
	\minipage{0.5\textwidth}
		\includegraphics[width=\linewidth]{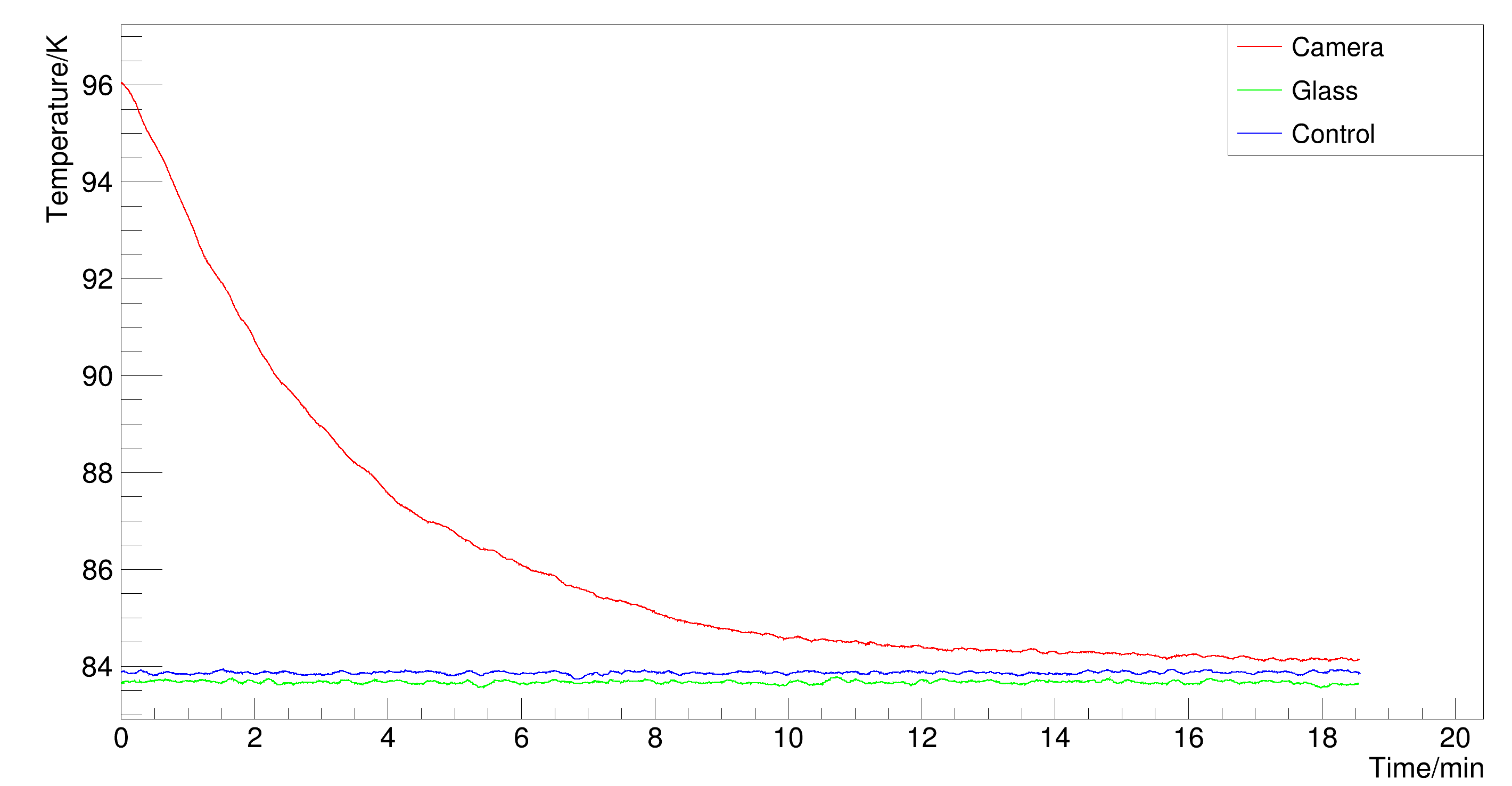}
	\endminipage\hfill
	\label{temprise}
	\caption{Local temperatures inside the camera module (red), outside the camera module (green) and in the dewar (blue) during camera operation. Left, the temperature change following camera switch on, right, the temperature change following camera switch off.\label{camonoff}}
\end{figure}

The cameras dissipate power when they are operational, and therefore produce heat.  The extent of this self heating was measured for a camera inside a stainless steel (SS) housing module (as described in section \ref{35t}). PT100 temperature sensors were placed inside the housing, on the outside of the housing, and in the cryogenic liquid bath, disconnected from the camera system.  For a liquid argon ambient temperature of 84$\pm$3K, the local temperature inside the housing was found to reach equilibrium at 96$\pm$3K, after 15 minutes.  In this time, the temperature at the outside of the housing does not measurably increase, or differ from the control measurement.  When the cameras are switched off, the temperature inside the housing reaches equilibrium with the outside temperature after around 15 minutes, as shown in figure \ref{camonoff}

During the camera testing phase, some inconsistency was observed between cameras in the reliability of power cycling at cryogenic temperatures.  Whilst all of the batch of 20 cameras operated reliably when cooled down when powered on, only 10 were operated from cold in a consistently reliable way.  From the cameras which operated less reliably in the cold, 4 were unable to switch on in the cold, 6 were able to switch on at a slightly raised local temperature.  The variability in behaviour of the cameras is due to operation outside of the rated temperature specification of the device. 

For operation over a lengthy period in the cold, it is necessary that all of the cameras are able to switch on at cryogenic temperatures, so that they remain functional in the instance that there is a power failure. All of the cameras selected for further use were able to power cycle in the cold, however for the longer term operation in the cold it was decided to implement a failsafe mechanism which allows local heating at the cameras.

\begin{figure}
\centering
\includegraphics[width=0.6\linewidth]{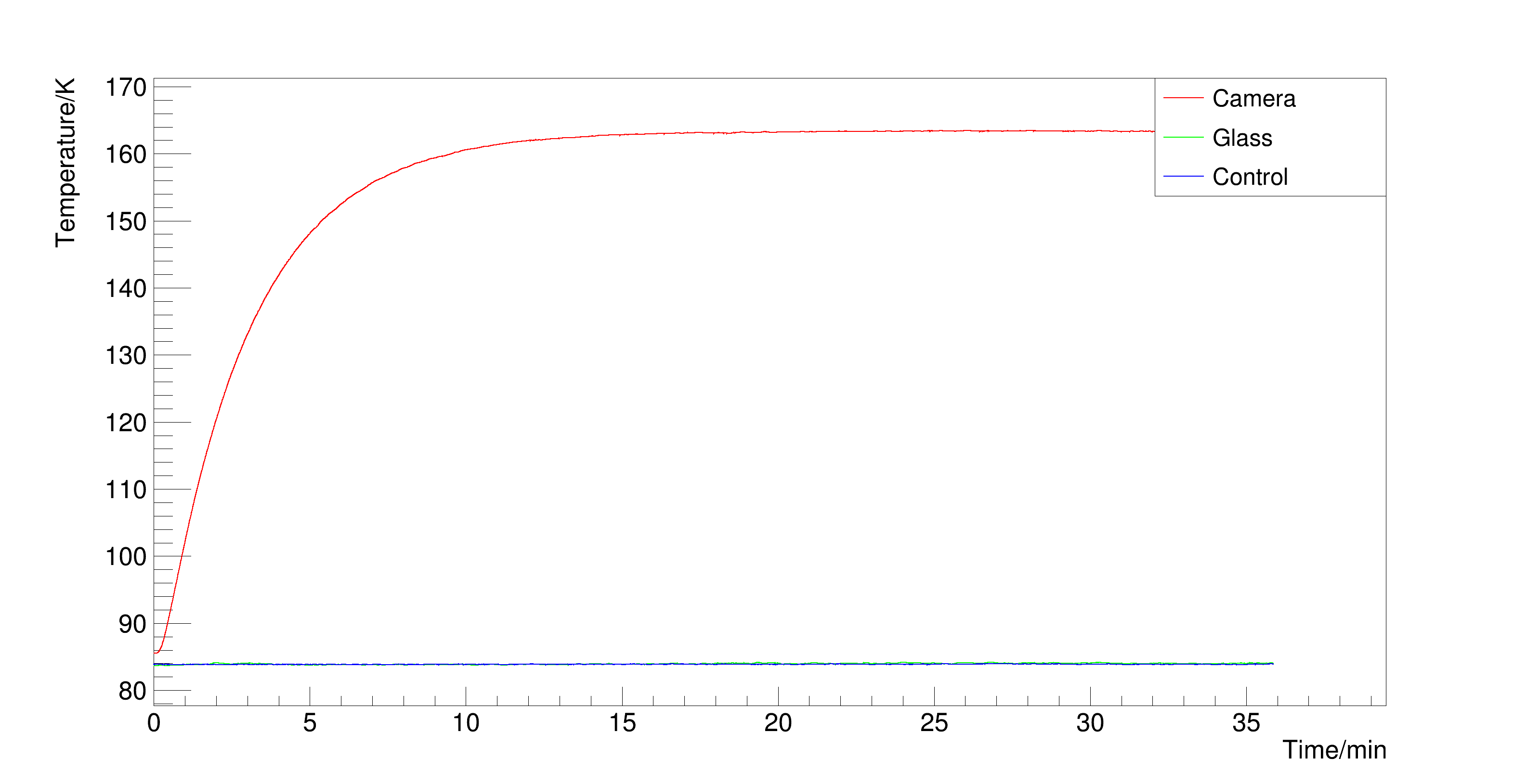}
\caption{Local temperatures inside the camera module (red), outside the camera module (green) and in the dewar (blue) with heater operational.  This shows the equilibrium temperature which is reached inside the module where the heater is on. \label{heater}}
\end{figure}

A camera module (as described in section \ref{module}) was designed as an environment in which local heating could take place.  The heating system consisted of a pair of thick film power resistors of 20$\Omega$ connected in series, mounted on either side of a camera. The resistors were powered with 10V, 250$\,$mA, giving a maximum power output of 2.5$\,$W. With the camera module submerged in liquid nitrogen, the heating effect is shown in figure \ref{heater}.  An equilibrium temperature of 163.4$\,$K was reached after the heating resistors have been on for 17 minutes.  

\section{Cameras in DUNE 35t Prototype}\label{35t}

The DUNE 35-ton prototype consists of a LAr TPC inside a membrane cryostat built at Fermilab in 2015. The purpose was to test various engineering design elements of the full DUNE far-detector. One of the goals was to test the stability of the HV system design against breakdown in argon. To this end, a system of six cameras for visual monitoring of sparks or coronae was installed in various high-field locations inside the cryostat, including the cathode plane, HV feedthrough, and field cage. As described in section \ref{campar}, the cameras have been demonstrated to be sensitive to the light emitted during one of these breakdown events, have the timing capability to trigger on and record them, have the spatial resolution to locate the source of the breakdown, and are functional when cooled to 83K. All of these capabilities make the system useful in diagnosing design flaws which are unable to be determined using other breakdown monitoring methods, such as measuring power supply current spikes.

In addition to monitoring HV breakdown, two cameras were installed in order to diagnose possible cryogenic systems issues in the cooling sprayers and the phase separator after the cryostat was fully sealed and visual monitoring was otherwise impossible.

\begin{figure}
	\minipage{0.25\textwidth}
		\includegraphics[width=\linewidth]{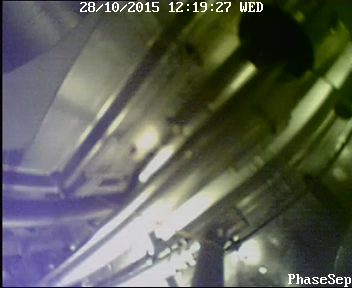}
	\endminipage\hfill
	\minipage{0.25\textwidth}
		\includegraphics[width=\linewidth]{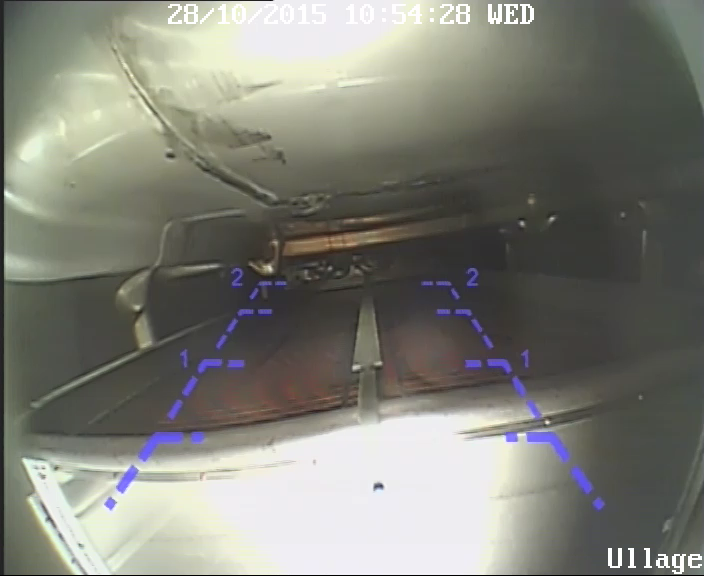}
	\endminipage\hfill
	\minipage{0.25\textwidth}
		\includegraphics[width=\linewidth]{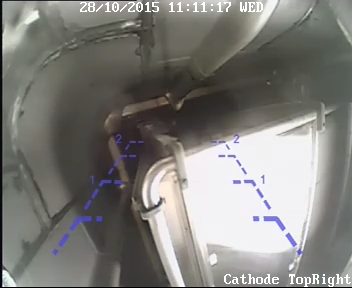}
	\endminipage\hfill
	\minipage{0.25\textwidth}
		\includegraphics[width=\linewidth]{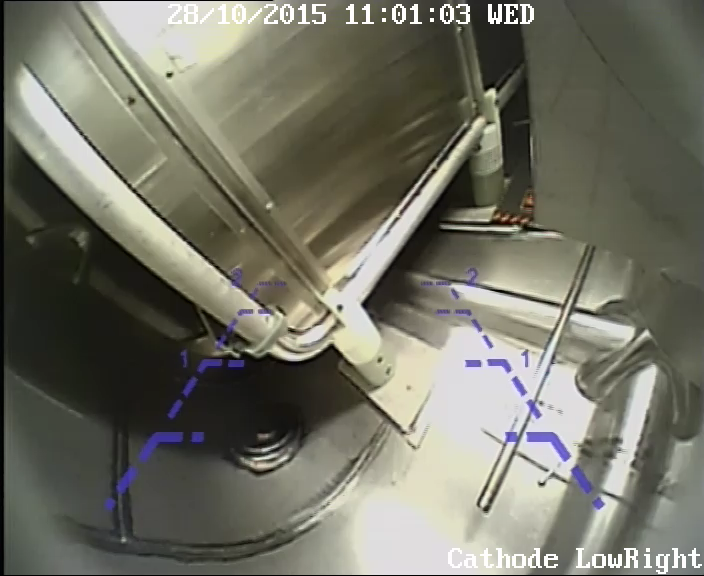}
	\endminipage\hfill

	\minipage{0.25\textwidth}
		\includegraphics[width=\linewidth]{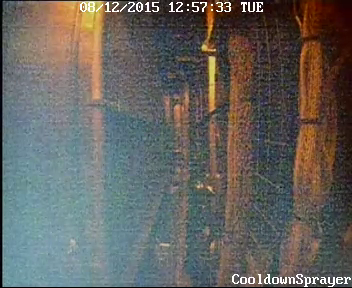}
	\endminipage\hfill
	\minipage{0.25\textwidth}
		\includegraphics[width=\linewidth]{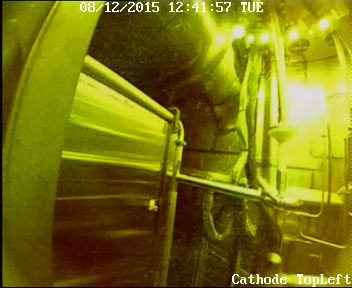}
	\endminipage\hfill
	\minipage{0.25\textwidth}
		\includegraphics[width=\linewidth]{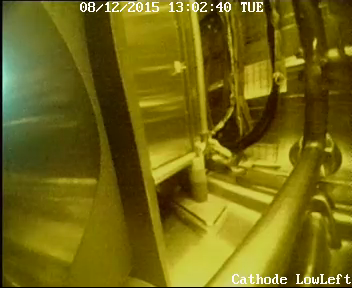}
	\endminipage\hfill
	\minipage{0.25\textwidth}
		\includegraphics[width=\linewidth]{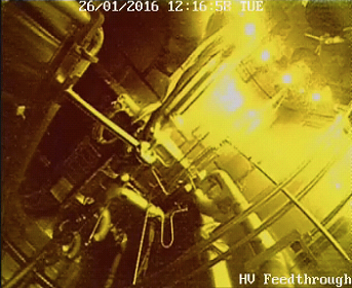}
	\endminipage\hfill

\caption{The calibration images for the 8 cameras in the system.  Upper (left to right): Phase separator, ullage, cathode top right, bottom right. Lower (L to R) cooldown sprayers, cathode top left, bottom left,  and high voltage feedthrough.  The upper images were taken with a halogen light illuminating the cryostat, prior to it being sealed up.  The lower images were taken with the LED ring light on, with the cryostat sealed up.  All images are left-right inverted due to software.
\label{cryostatpics}}
\end{figure}

Fields-of-view of each of the eight cameras are shown in figure \ref{cryostatpics}. The modules, mounts, and operational system are described below.

\subsection{Camera module}\label{module}

\begin{figure}
\centering
\includegraphics[width=0.8\textwidth]{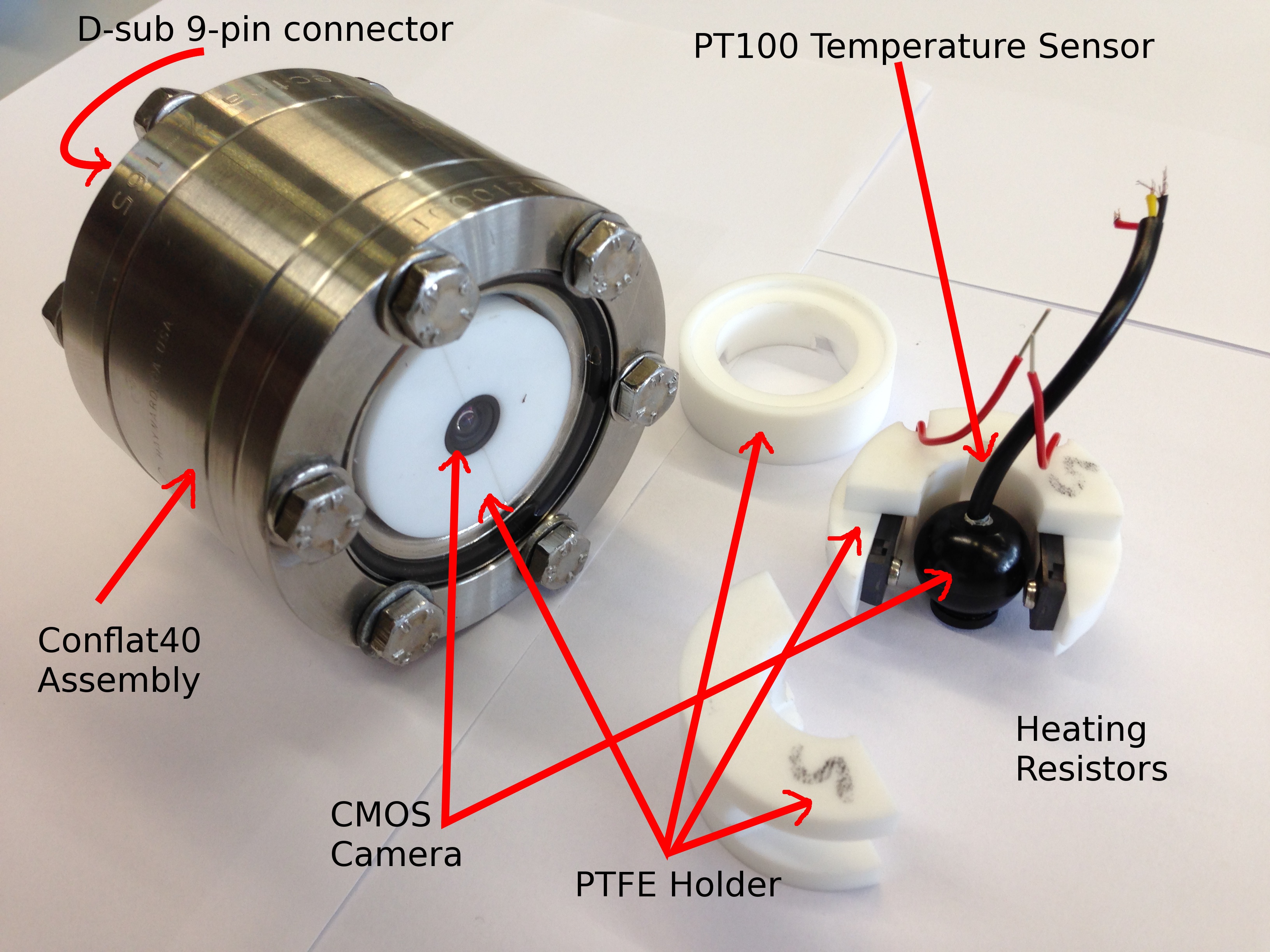}
\caption{A sealed camera module, and the components of a module.\label{modulepic}}
\end{figure}

\begin{figure}
\centering
\includegraphics[width=0.6\textwidth]{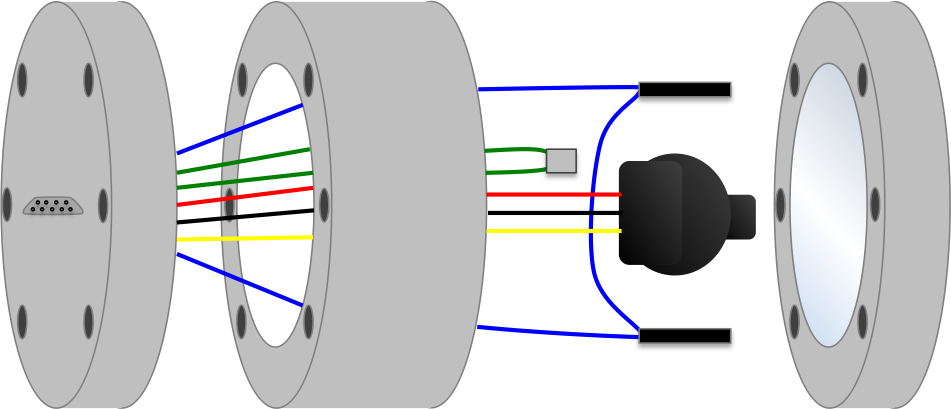}
\caption{From left to right: CF40 flange with 9-pin D-sub feedthrough, double sided CF40 flange, PT100 sensor (green wires), camera in centre (red, black, yellow wires), two heating resistors (blue wires) on either side of the camera connected in series, optical viewport on CF40 flange.\label{cameraschem}}
\end{figure}

Each camera is housed in a self contained module, which prevents any contamination of the liquid argon from the camera material, and provides an extremely robust and easily installable system. Each module contains a CMOS camera, PT100 temperature sensor and a pair of 20$\Omega$ heating resistors, which are held in place by a custom made PTFE liner. A photograph of this is shown in figure \ref{modulepic} and a schematic in figure \ref{cameraschem}.

The camera module housing is  composed of a double sided CF40 conflat flange, with a CF40 D-subminiature 9 pin plug and a CF40 Kodial glass optical viewport.  Each module has a diameter of 70$\,$mm and a length of 55$\,$mm, and has mass of approximately 1.2$\,$kg.  The PTFE liner fits snugly to the flange and components, holding the camera lens in contact with the glass viewport.  The heating resistors flank the camera, and the temperature sensor is mounted directly behind the camera.  

The heating resistors, as described in section \ref{campar}, were included as a failsafe.  They give the option to raise the local temperature of the camera, but are unnecessary for camera operation.  The PT100 temperature sensor, with precision $\pm$0.5$^{\circ}$C is used to monitor the operating temperature of the camera, and the local temperature when operating the heating resistors.  Each module was assembled and sealed in a helium environment, and He leak checked for tightness to ensure a vacuum seal.  

\subsection{Camera system}

\begin{figure}
\centering
\includegraphics[width=0.6\textwidth]{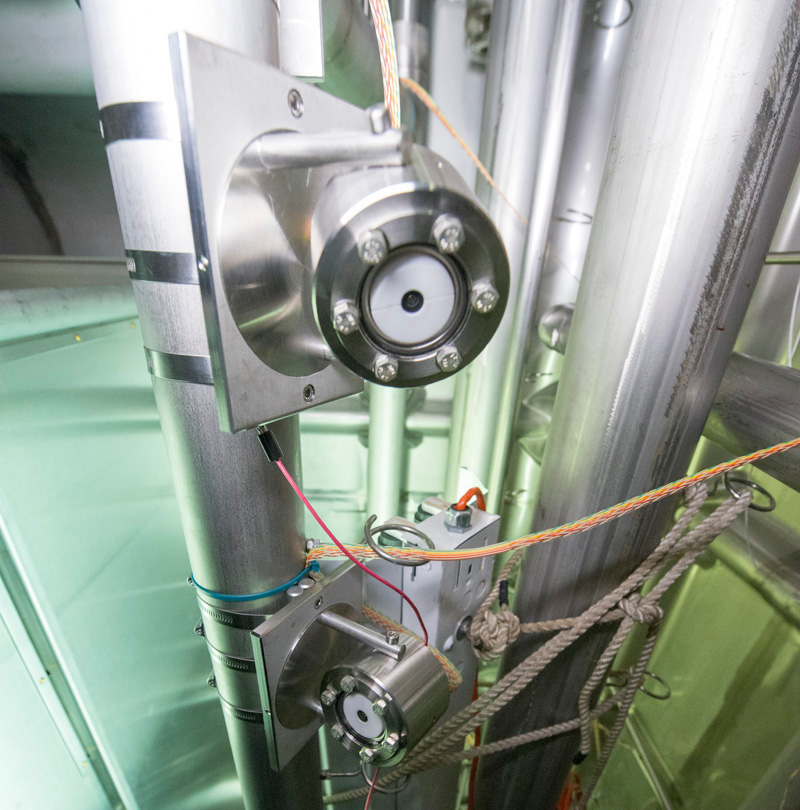}
\caption{Two camera modules and mounts clamped to 3" SCH cryogenic pipes in the 35t cryostat. \label{mountin35t}}
\end{figure}

The camera modules were individually mounted in the cryostat using a custom designed mounting bracket, as shown in figure \ref{mountin35t}.  These mounts were designed to attach to existing cryogenic pipework, and with the flexibility to be rotated along 3 axes, in order to select and fine tune the orientation of the camera prior to fixing the positions when in the cryostat.  

\begin{figure}
\includegraphics[width=\linewidth]{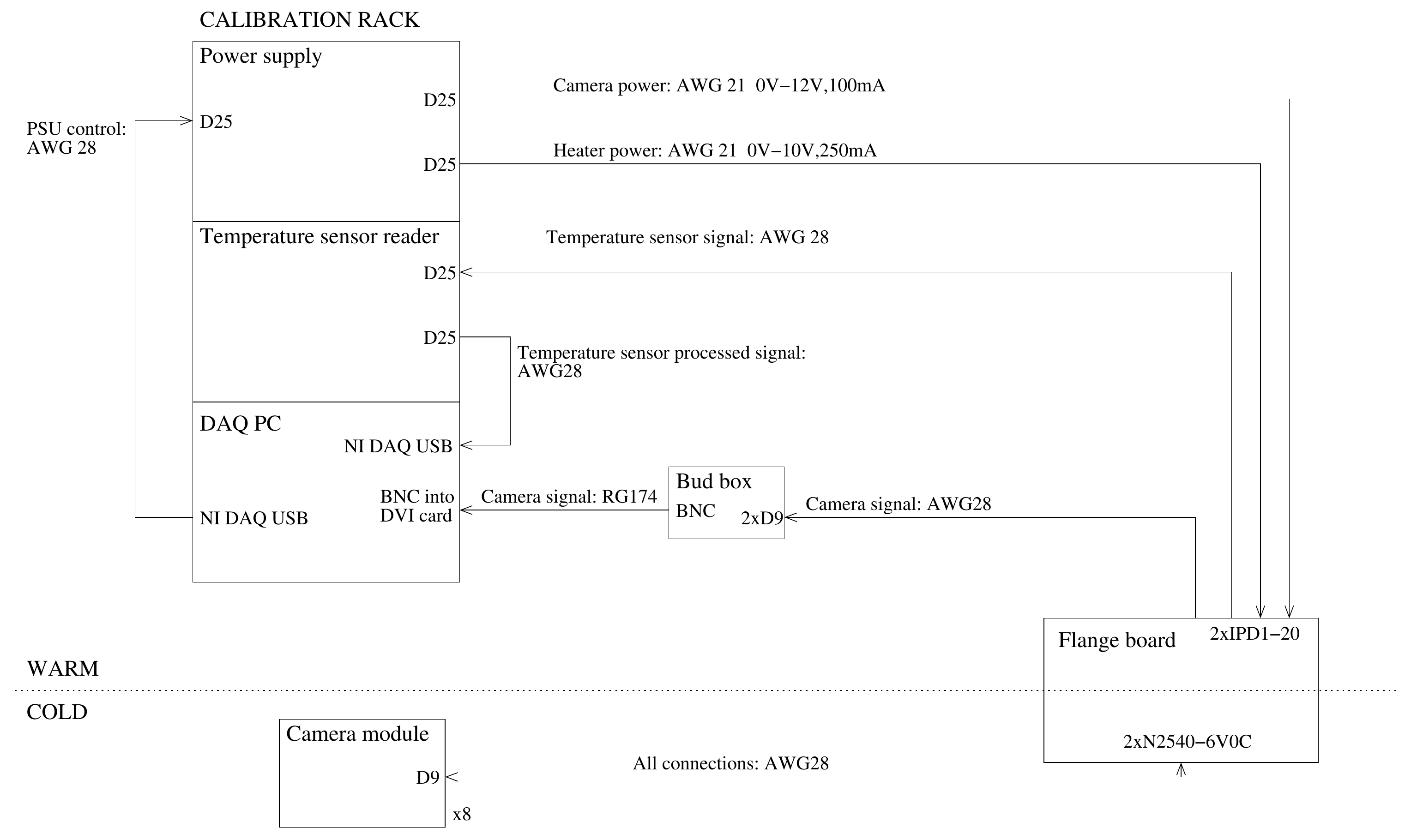}
\caption{Full system block diagram for the camera modules in the DUNE 35t Prototype.\label{blockdiag}}
\end{figure}

Data acquisition, operation and control of the camera modules was via a rack-based system comprising power supplies, temperature sensor reader, and data acquisition (DAQ) and control computer system.  These are connected to the 8 camera modules via the flange board, a printed circuit board, which takes signals from the majority of the systems within the cryostat and connects them to the DAQ on the warm side of the cryostat.  The system and its interconnects are detailed in figure \ref{blockdiag}.

A NI LabVIEW \cite{LabVIEW} program was used to send operation control signals to the power supplies, and monitors temperature inputs, via a NI USB-6009 device.  The power supply is a custom made, remotely controllable device, with outputs of 12$\,$V and 10$\,$V DC for the cameras and heating resistors respectively.  The channels are individually controlled using digital logic signals sent from the NI USB-6009.  The raw signal from the PT100s is signal processed by a custom made device, which outputs a voltage signal which is read into the computer via the NI USB-6009.  Off-site control of these systems is possible using the Remote Desktop system, over the Fermilab network.

The camera signals were read out by a digital video recorder (DVR), which is remotely accessible using SwannView Link \cite{DVR}, a commercially available surveillance programme.  This software is configured to record when the software detects movement between successive frames, as described in section \ref{campar}. This video, titled by timestamp is saved in a file on the DVR hard drive.  A live video feed is also available, which was used when monitoring the cooldown process, and also when ramping up the high voltage.

\subsection{Camera performance}

Post installation in the 35t cryostat, the camera system was characterised at room temperature.  The software trigger system was tested, and successfully triggered on a pulsing Xe flashlamp which forms part of the purity monitoring system.  During cryostat filling, the camera system was recording continuous video footage, with the LED ring illuminating the cryostat.  This was used to visually monitor the liquid level.  Footage taken of the liquid level passing a camera is shown at reference \cite{paperlink}.  


The cameras were successfully operational at cryogenic temperatures throughout the 10 weeks of Phase II of the 35t cooldown.  Over the course of the run, the system was power cycled successfully three times, with the duration ranging from 30 minutes to 9 days. These outages were necessary for TPC noise troubleshooting, and due to a power outage, rather than any camera system component failure.

\subsubsection{Degradation over time}

The cameras operated in the 35t show some degradation in the picture quality and resolution in the cold.  The authors suggest this is for multiple reasons; the signal transmission length, power cycling, and prolonged exposure to the cold being the most significant.  In the testing phase described in section \ref{campar}, the chrominance of the video signal was shown to change, but the resolution remained the same.  

Over the duration of the cold operation, the picture quality of the cameras changed in a number of ways.  With the cryostat in darkness, there were a greater number of saturated and flickering pixels.  With the cryostat lit by the LED ring, the colour depth decreased, and noise increased, producing a more pixelated image. The resolution consequently decreased, although the differences in resolution across the 10 week cooldown period are minimal compared to the difference in resolution between warm and cold.  There is significant variation between cameras in terms of degradation. 

\begin{figure}
\centering
\includegraphics[width=0.8\textwidth]{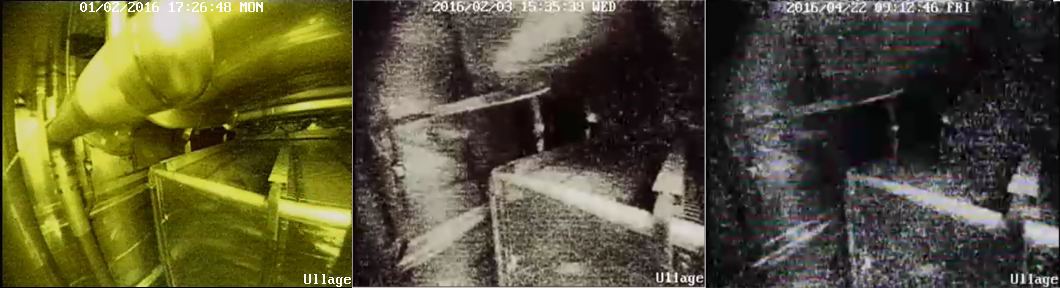}
\includegraphics[width=0.8\textwidth]{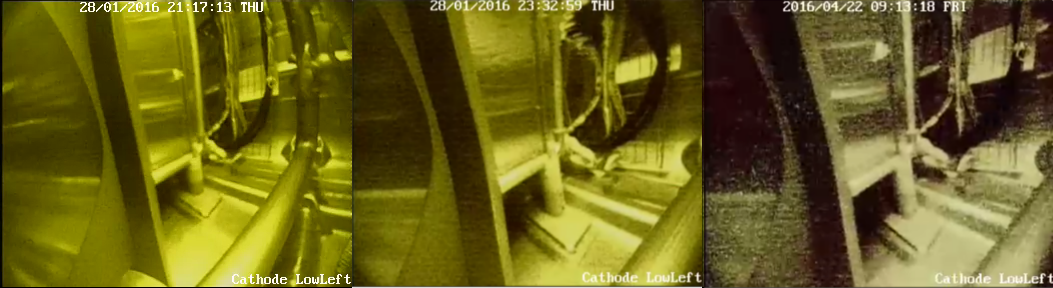}
\caption{The variation in picture quality degradation is illustrated by the changes in camera 1 (upper) and camera 4 (lower) over time.  Left: prior to cooldown, Centre: Immediately post-cooldown, Right: after 10 weeks submerged in LAr.  The field of view changes due to the change in refractive index. Note that these are full colour images as recorded by the DVR with no post-processing.}
\label{camndeg} 
\end{figure}

Figure \ref{camndeg} shows images taken over the course of the run in cameras observing the Ullage and Cathode bottom left section respectively.  The left most images show the picture prior to cooldown, the 2 images to the right show images taken under liquid argon, and consequently have a smaller field of view, due to the relative refractive indices of liquid and gaseous argon.  The second image shows the camera picture immediately post liquid fill.  Whilst four of the cameras show little difference between their images pre/post filling, four cameras have a significantly degraded image quality.  This variation in behaviour is somewhat expected, due to the variation in the cameras in the testing phase.  The small amount of degradation which occurs over the course of usage in the cold is predominantly an increase in noise, rather than increased granularity, hence the change in resolution is minimal. 

\subsubsection{High voltage monitoring}

The voltage on the cathode in the 35t cryostat operated stably at 60$\,$kV for several weeks, with no indication of high voltage breakdown from either the power supply, the TPC itself or the high voltage monitoring cameras.  There were two suspected breakdowns at this voltage, but the cameras were non-operational due to a power outage.  In this instance, a breakdown is defined as an increase in current drawn from the power supply to above the threshold of 8$\,\mu$A, which caused the power supply to trip.

An increase of the cathode HV to 135$\,$kV in low-purity argon led to a total of four breakdowns, three of which were detected by the camera system. The camera system triggered on the breakdowns, writing video data to disk, however, the images recorded do not give a clear picture of the position of a source of light.  In the case of some of the cameras, the final breakdown at 135kV triggered on electrical interference with the camera system, caused by the power surge.  This breakdown also caused other electrical systems within the cryostat to trip. 

There are several possible reasons the camera system was not able to pinpoint the location of the breakdowns: the breakdown didn't take place in the Cryostat nor not in the field of view of any of the cameras, the breakdown was either too fast or of not great enough intensity for the cameras to detect the sparks, or the triggering system was not appropriate due to the degradation in picture quality and thus the sensitivity and efficiency of the trigger system were compromised

Although the cameras were able to detect sparks in a bench-top cryogenic test stand, they did not observe high voltage breakdown in the 35t. Therefore the camera system has not been conclusively demonstrated to have the sensitivity for high voltage monitoring over larger distances. The cameras have nevertheless proven extremely useful for monitoring purposes in the 35t. 


\section{Conclusions}

A CMOS camera system has been successfully designed, constructed, characterised and operated at cryogenic temperatures.  This represents a new method of monitoring the inside of large cryogenic detectors.  Eight camera modules were installed, calibrated, and operated in the DUNE 35t prototype, and there are plans to install similar cameras in future LAr TPCs such as SBND \cite{SBND}.  

Four of the camera modules with best picture quality will be used in future high voltage tests for DUNE in the 35t cryostat.  It is anticipated that this test will lead to an improved trigger system for the cameras, which will improve the efficiency of the system and therefore test definitively if the cameras have the sensitivity to be appropriately used for monitoring high voltage breakdown.

The camera system has been demonstrated to be an excellent method for reliably monitoring the status of the inside of a cryostat, especially for monitoring the liquid level relative to the TPC, and observing cooldown.  This work paves the way for future work using cryogenic cameras inside the cryostat.  With suitable effort this system can be optimised to have improved characteristics and stability in liquid argon.

\section{Acknowledgements}
The authors like to acknowledge the support for this work though the STFC grant award ST/N000277/1 and associated awards from STFC for DUNE-UK.

We would also like to acknowledge the invaluable help of the following individuals: 
Trevor Gamble and University of Sheffield Physics Department Mechanical Workshop for the design and manufacture of the camera mounting system, and camera liner system.  Bob Bridgeland of University of Warwick Physics Department Electronics Workshop for design and manufacture of power supply and temperature sensor reader units.  Linda Bagby for help and support with the rack and electronics infrastructure.  Lee Scott for mounting and fine-tuning the positions of the cameras in the cryostat.  Alan Hahn for his help with 35t operations.

\end{document}